\documentstyle[12pt,epsfig]{article}
\begin{document}
\centerline{\Large\bf Limits on the Time Evolution of Space}
\centerline{\Large\bf Dimensions from Newton's Constant}
\vspace*{0.050truein}
\centerline{Forough Nasseri\footnote{Email: nasseri@fastmail.fm}}
\centerline{\it Physics Department, Sabzevar University of
Tarbiat Moallem, P.O.Box 397, Sabzevar, Iran}
\centerline{\it Khayyam Planetarium, P.O.Box 769, Neishabour, Iran}
\begin{center}
(\today)
\end{center}

\begin{abstract}
Limits are imposed upon the possible rate of change of 
extra spatial dimensions in a decrumpling model Universe with
time variable spatial dimensions (TVSD) by considering the time variation
of (1+3)-dimensional Newton's constant. Previous studies on the time
variation of (1+3)-dimensional Newton's constant in TVSD theory had not
been included the effects of the volume of the extra dimensions
and the effects of the surface area of the unit sphere in D-space
dimensions. Our main result is that the absolute value of the present
rate of change of spatial dimensions to be less than
about $10^{-14} {\rm yr}^{-1}$. Our results would appear to provide a
{\it prima facie} case for ruling the TVSD model out. We show that
based on observational bounds on the present-day variation of Newton's
constant, one would have to conclude that the spatial dimension of the
Universe when the Universe was ``at the Planck scale'' to be less than or
equal to $3.09$. If the dimension of space when the Universe was ``at the
Planck scale'' is constrained to be fractional and very close to $3$,
then the whole edifice of TVSD model loses credibility.
\end{abstract}

\date{today}

\section{Introduction}
Although time variability of spatial dimensions have not been firmly
achieved in experiments and theories, such dynamical behavior of the
spatial dimensions should not be ruled out in the context of cosmology
and astroparticle physics.

This article studies the time variation of the (1+3)-dimensional
Newton's constant in a model Universe with time variable space dimension
(TVSD) following the original idea presented in ``A model Universe
with variable dimension: expansion as decrumpling'' \cite{r101}.

Decrumpling model of the Universe is a new sort of cosmological
scenario based on the assumption that the basic building blocks of
the spacetime are fractally structured \cite{r101}-\cite{r107}.
In the original papers \cite{r101}, the spatial dimension
of the Universe was considered as a continuous time dependent variable.
As the Universe expands, its spatial dimension decreases continuously,
thereby generating what has been named a decrumpling Universe. Then
this model has been overlooked and the quantum cosmological aspects,
as well as, a possible test theory for studying time evolution of Newton's
constant have also been discussed \cite{r102,r103}. Chaotic inflation in
TVSD theory and its dynamical solutions have also been studied in Refs.
\cite{r104,r105}.

The idea of changing the spatial dimension of the Universe dynamically
as has been suggested in the original papers \cite{r101} is a bold one
and perhaps seems to be unique and novel but we believe that daring and
speculative ideas like this should be explored.

There have been attempts to identify a fractal dimension for the matter
distribution in space using either cosmic microwave background radiation
(CMBR) or galaxy distribution \cite{4mn,5mn}. Aside from the actual
dimension of space or the matter distribution in it, it is interesting
to study the cosmological consequences of a fractal and variable space
dimension. All critiques of space dimensionality other than $3$ rely upon
cosmologically small scale observations \cite{6mn}. Therefore, one could
ask about the consequences of a dynamical space dimension in cosmological
time and space scales. A proposed way of handling such a concept is using
the idea of decrumpling coming from polymer physics \cite{r204,r205,r206}.
In recent years, the physics community has witnessed a spectacular
revival in interest for the evolution of extra spatial dimensions
\cite{r302,john,barrow,r108,ann,mich,tors,gu,r203}.
The topic of the fractal dimension has also been
studies in Refs. \cite{fed,thd,tsm,otn,hill,carl,mittal,lya}.
The evolution of the fractal dimension of a self-similar Universe in the
context of Newton's gravitation has been discussed in Ref. \cite{abda}.

Here, we will be concerned with the approaches proposed in Ref.
\cite{r101} where the cosmic expansion of the Universe is named
decrumpling expansion and is due to the decrease of the spatial dimensions.
The most important difference between TVSD model/theory and other
attempts about the evolution of the spatial dimension is that in this
theory the number of the extra spatial dimensions changes with time while
in other theories the size of the extra spatial dimension is a dynamical
parameter, see Refs. \cite{r302,john}.

Another subject which lately has attracted much attention is the time
variation of the physical constants for example the fine structure
constant and the Newton's gravitational constant
(see Ref. \cite{r304} for a thorough review or Refs. \cite{sdegl,fgb}
for a brief study).

In the previous studies Refs. \cite{r103,ali} about the effective time
variation of Newton's constant in TVSD model the effects of the surface
area of the unit sphere in $D$-space dimensions had not been included
and the effects of the volume of the extra spatial dimensions had
been included for closed Universe which is wrong because we know based on
recent observational data that our Universe is flat.
Here, our study will include correctly these important effects and will
give a lower bound about $10^{-14} {\rm yr}^{-1}$ on the rate of the
change of the spatial dimension based on the time variation of
($1+3$)-dimensional Newton's constant. We use the relationship between
Newton's constant in ($1+3$) and ($1+D$)-dimensional theories as
obtained by Gauss' law in Ref. \cite{r301}.

In this paper we will answer to this main question can such models be ruled
out observationally? Our results would appear to do so. In particular, our
conclusion is that based on observational bounds on the present-day
variation of Newton's constant, one would have to conclude that the space
dimension when the Universe was ``at the Planck scale'' to be less than or
equal to $3.09$. If the dimension of space when the Universe was ``at the
Planck scale'' is constrained to be fractional and very close to $3$,
then the whole edifice of this model loses credibility.

We will use a natural unit system that sets $k_B$, $c$ and $\hbar$
all equal to $1$, so that $\ell_P=M_P^{-1}=\sqrt{G}$.
To read easily this article we also use the notation $D_t$ instead of
$D(t)$ that means the space dimension $D$ is as a function of cosmic
time.

The plan of this article is as follows. Since TVSD model/theory
is so far from the mainstream of current researches, in section 2
we will give a brief review of the idea of TVSD theory and of its
physical content presented in Refs. \cite{r101}-\cite{r107}.
In section 3, we will confront this idea with the time variation of
Newton's constant, showing that the effect of the total volume of
the extra spatial dimension and the surface area of the unit sphere in
$D$-space dimension are important while in previous studies in
Ref. \cite{r103} have not been included. In section 4, we explain that
our results would appear to provide a {\it prima facie} case for ruling
the TVSD model out and we also study the
time variation of the spatial dimension from viewpoint of the anthropic
cosmological principle. Finally, we discuss our results and conclude in
section 5.

\section{Review of TVSD Theory}
\subsection{Motivation for choosing a Universe with TVSD}
\noindent
Problems of the standard model and its difficulties with the concept of
quantum gravity and the early Universe at the Planck time provide
us enough reasons to look for viable model Universes.
Moreover, the ongoing experiments related to CMB will provide
us a wealth of data suitable to test all the theories of spacetime
and gravity. Even the act of verifying cosmological models based on
general relativity needs looking for viable theories differing
from it to see the degree of its testability and viability.
These are the main reasons we are studying decrumpling Universe
based on time variable space dimensions. It has been shown in
Ref. \cite{r102} that this idea can be implemented successfully in a
gravitational theory and cosmological model based on it. The free
parameter of the theory may then be fixed by observational data
\cite{r103}.

The idea of having spacetime dimensions other than $1+3$ goes
back to Kaluza-Klein theory. The generalization of this concept to
string theories with space dimension more than three, but still an
integer, and a constant is well known \cite{{r201},{r202}}. This,
being considered for the high energy limit in the Universe or for
the dimension of space at the Planck time, has encouraged people
to suggest that the dimension of space in the lower energy limit,
or for the actual structured Universe, be other than three.

A possible time dependence of spacetime dimensionality is derived
as an effect of entropy conservation in Ref. \cite{r203}. There it
turns out that, going back in time, the dimension increases first
very slowly up to about the Planck time, and increases more
rapidly thereafter. This generic trend is insensitive to the
assigned entropy value. In fact a minimum value for the size of
the Universe, being about the Planckian size, is also obtained.
The dynamical model we are going to follow has the same time
behavior for the space dimension. A minimum size for the Universe
is also built into our dynamical model.

Our treatment in this paper is based on a cosmological model,
where the number of spatial dimensions decreases continuously as the
Universe expands, presented in the pioneer works \cite{r101}.
A proposed way of handling such a concept is using the idea of decrumpling
coming from polymer physics \cite{{r204},{r205},{r206}}.
In this model the fundamental building blocks of
the Universe are like cells being arbitrary dimensions and having, in
each dimension, a characteristic size $\delta$ which maybe of the order of
the Planck length ${\mathcal O}$($10^{-33}$ cm) or even smaller.
These ``space cells'' are embedded in a ${\mathcal D}$ space, where
${\mathcal D}$ may be up to infinity. 
Therefore, the space dimension of the
Universe depends on how these fundamental cells are configured in this
embedding space. The Universe may have begun from a very crumpled
state having a very high dimension ${\mathcal{D}}$ and a size $\delta$,
then have lost dimension through a uniform decrumpling which we see
like a uniform expansion. The expansion of space, being now understood like
a decrumpling of cosmic space, reduces the spacetime dimension continuously
from ${\mathcal{D}}+1$ to the present value $D_0+1$.
In this picture, the Universe can have any space dimension.
As it expands, the number of spatial dimensions decreases continuously.
The physical process that causes or necessitates such a decrease in
the number of spatial dimensions comes from how these fundamental cells
are embedded in a ${\mathcal{D}}$ space.

As an example, take a limited number of small three-dimensional beads.
Depending on how these beads are embedded in space they can configure to a
one-dimensional string, two-dimensional sheet, or three-dimensional sphere.
This is the picture we are familiar with from the concept of crumpling
in polymer physics where a crumpled polymer has a dimension more than 1.
Or take the picture of a clay which can be like a
three-dimensional sphere, or a two-dimensional sheet, or even a
one-dimensional string, a picture based on the theory of fluid membranes.

The major formal difficulty to implement this idea in a
spacetime theory with variable space dimension is that the
measure of the integral of the action is variable and therefore
some part of integrand. However, taking into account the
cosmological principle, i.e. the homogeneity and isotropy of
space, the formulations are simplified substantially. It then becomes
possible to formulate a Lagrangian for the theory and write down the
corresponding field equations. This Lagrangian is however 
not unique \cite{r102}.

The original decrumpling model of the Universe seems to be
singularity free, having two turning points for the space
dimension \cite{r101}. The authors of Ref. \cite{r106}
criticize the way of generalizing the standard cosmological model
to arbitrary variable space dimension used in Ref. \cite{r101} and propose another
way of writing the field equations. Their model shows no upper
bound for the dimension of space, see also Ref. \cite{r107}.

Later on this scenario was
extended to the class of multidimensional cosmological models,
where extra factor spaces play the role of the matter fields. In
this multidimensional cosmological model an inflationary solution
was found together with the prediction that the Universe starts
from a nonsingular spacetime \cite{r108}.
Moreover, the dynamics of the dimensions in
factor space cosmology has been studied in Ref. \cite{r108}.

A new way to generalize the gravitational action in constant
dimension to the case of dynamical dimension is 
proposed in Ref. \cite{r102}. 
There, it is shown that the generalization of the gravitational 
action to the dynamical
dimension is not unique. Moreover, in contrast to the earlier
works in Ref. \cite{r101}, the dependence of the
measure of the action on space dimension is taken into account.
This new decrumpling model is studied in detail in Ref. \cite{r102}.
The generalization of the action, the Lagrangian, the equations of
motion to dynamical space dimension, the time evolution of the
spatial dimension, numerical results for the turning points of the model,
and its quantum cosmology within the concept of the Wheeler-DeWitt
equation are derived. It is shown that the corresponding potential of
the model has completely different behavior from the potential of the
de Sitter minisuperspace in three-space. Imposing the appropriate boundary
condition in the limit $ a \to +\infty$, and using the
semiclassical approximation, the wave function of the model is
also obtained. It is then seen that in the limit of constant space
dimension, the wave function is not well-defined. It can approach
to the Hartle-Hawking wave function or to the modified Linde wave
function, but not to that of Vilenkin. In the limit of constant
spatial dimension, the probability density approaches to Vilenkin,
Linde and others' proposal; i.e. to the probability
density ${\mathcal{P}} \propto \exp(2S_E)$, or more generally
$\exp(-2|S_E|)$, where $S_E$ is the Euclidean action of the
classical instanton solution.
Chaotic inflation in TVSD scenario has been studied in Ref. \cite{r104}
and its solutions are presented in Ref. \cite{r105}.
In Ref. \cite{r103} the effective time variation of Newton's constant
in TVSD theory has also been investigated.

\subsection{Relation between the effective space dimension
$D(t)$ and characteristic size of the Universe $a(t)$}
\noindent
Assume the Universe consists of a fixed number $N$ of universal
cells having a characteristic length $\delta$ in each of their
dimensions. The volume of the Universe at the time $t$ depends
on the configuration of the cells. It is easily seen that
\cite{r101}
\begin{equation}
{\rm vol}_D({\rm cell})={\rm vol}_{D_0}({\rm cell})\delta^{D-D_0}.
\end{equation}

Interpreting the radius of the Universe, $a$, as the radius of
gyration of a crumpled ``universal surface''\cite{r206},
the volume of space can be written \cite{r101}
\begin{eqnarray}
a^D&=&N {\rm vol}_D({\rm cell})\nonumber\\
   &=&N {\rm vol}_{D_0}({\rm cell}) \delta^{D-D_0}\nonumber\\
   &=&{a_0}^{D_0} \delta^{D-D_0}
\end{eqnarray}
or
\begin{equation}
\label{31}
\left( \frac{a}{\delta} \right)^D=
\left( \frac{a_0}{\delta} \right)^{D_0} = e^C,
\end{equation}
where $C$ is a universal positive constant. Its value has a strong
influence on the dynamics of spacetime, for example on the dimension
of space, say, at the Planck time. Hence, it has physical and cosmological
consequences and may be determined by observations. The zero subscript in any
quantity, e.g. in $a_0$ and $D_0$, denotes its present values.
We coin the above relation as a``dimensional constraint" which relates
the ``scale factor" of our model Universe to the space dimension.
In our formulation, we consider the comoving length of the Hubble radius
at present time to be equal to one. So the interpretation of the scale
factor as a physical length is valid.
The dimensional constraint can be written in this form
\begin{equation}
\label{32}
\frac{1}{D}=\frac{1}{C}\ln \left( \frac{a}{a_0} \right) + \frac{1}{D_0}.
\end{equation}

It is seen that by expansion of the Universe, the space
dimension decreases. Note that in Eqs.(\ref{31}) and (\ref{32}),
the space dimension is a function of cosmic time $t$.
Time derivative of Eqs.(\ref{31}) or (\ref{32}) leads to

\begin{equation}
\label{33}
\dot{D}=-\frac{D^2 \dot{a}}{Ca}.
\end{equation}
It can be easily shown that the case of constant space dimension
corresponds to when $C$ tends to infinity. In other words,
$C$ depends on the number of fundamental cells. For $C \to +\infty$,
the number of cells tends to infinity and $\delta\to 0$.
In this limit, the dependence between the space dimensions and
the radius of the Universe is removed, and consequently we
have a constant space dimension.

\subsection{Physical Meaning of $D_P$}
We define $D_P$ as the space dimension of the Universe when the
scale factor is equal to the Planck length $\ell_P$.
Taking $D_0=3$ and the scale of the Universe today to be the present
value of the Hubble radius $H_0^{-1}$ and the space dimension at the
Planck length to be $4, 10,$ or $25$, from Kaluza-Klein and superstring
theories, we can obtain from Eqs. (\ref{31}) and (\ref{32})
the corresponding value of $C$ and $\delta$:
\begin{eqnarray}
\frac{1}{D_P}&=& \frac{1}{C} \ln \bigg( \frac{\ell_P}{a_0}\bigg) +
\frac{1}{D_0}= \frac{1}{C} \ln \bigg( \frac{\ell_P}{H_0^{-1}}\bigg) +
\frac{1}{3},\\
\delta&=&a_0 e^{-C/D_0}=H_0^{-1} e^{-C/3}.
\end{eqnarray}
In Table 1, values of $C$, $\delta$ and also
$\dot D|_0$ for some interesting values of $D_P$ are
given\footnote{Our solar year (the time required
for Earth to travel once around the Sun) is $365.24219$ days,
http://www.mystro.com/leap.htm.}.
These values are calculated by
assuming $D_0=3$ and
$H_0^{-1}=3000 {h_0}^{-1} {\rm Mpc} = 9.2503 \times 10^{27} {h_0}^{-1} 
{\rm cm}$, where $h_0=0.68 \pm 0.15$.
Since the value of $C$ and $\delta$ are not very
sensitive to $h_0$ we take $h_0=1$.

\begin{table}
\caption{Values of $C$ and $\delta$ for some values of
$D_P$ \cite{r102}. Time variation of space dimension today
has also been calculated in terms of sec$^{-1}$ and
yr$^{-1}$.}
\begin{tabular}{ccccc} \\ \hline\hline 
$D_P$ & $C$   & $\delta$ (cm)   & $\dot D|_0$ (sec$^{-1}$)  & $\dot D|_0$ (yr$^{-1}$) \\ \hline\hline
$3$           & $ +\infty$         &  $0$           & $0$ & $0$ \\ \hline
$4$           & $1678.797$         &  $8.6158 \times 10^{-216}$  & $-1.7374 \times 10^{-20} h_0$ & $ -5.4827 \times 10^{-13} h_0$  \\ \hline
$10$          & $599.571$          &  $1.4771 \times 10^{-59}$  &  $-4.8648 \times 10^{-20} h_0$ & $ -1.5352 \times 10^{-12} h_0$  \\ \hline
$25$          & $476.931$          &  $8.3810 \times 10^{-42}$  & $-6.1158 \times 10^{-20} h_0$ & $-1.9299 \times 10^{-12} h_0$ \\ \hline
$+\infty$     & $419.699$          &  $\ell_P$  & $-6.9498 \times 10^{-20} h_0$ & $ -2.1931 \times 10^{-12}h_0$ \\ \hline\hline
\end{tabular}
\end{table}

\subsection{Lagrangian formulations of the model and field equations}
\noindent
Usually, we are accustomed to work with an integer number of
dimension, and therefore a non-integer total number of
spatial dimensions looks peculiar.
It is clear that for non-integer value of space dimensions, one cannot
define the metric tensor. To overcome this problem, we use a
gravitational theory based on Lagrangian formulations.
In Ref. \cite{r102}, some shortcomings of the original Lagrangian
formulation of the model proposed in \cite{r101} have been
shown, regarding the fields equations and their results.

Let us define the action of the model for the special
Friedmann-Robertson-Walker (FRW) metric in an arbitrary fixed space
dimension $D$, and then try to generalize it to variable dimension.
Now, take the metric in constant $D+1$ dimensions in the following
form
\begin{equation}
\label{metric}
ds^2 = -N^2(t)dt^2+a^2(t)d\Sigma_k^2,
\end{equation}
where $N(t)$ denotes the lapse function and $d\Sigma_k^2$ is the line
element for a D-manifold of constant curvature $k = + 1, 0, - 1$. The
Ricci scalar is given by
\begin{equation}
\label{r}
R=\frac{D}{N^2}\left\{\frac{2\ddot a}{a}+(D-1)\left[\left(\frac{\dot a}{a}
\right)^2 + \frac{N^2k}{a^2}\right]-\frac{2\dot a\dot N}{aN}\right\}.
\end{equation}
Substituting from Eq.(\ref{r}) in the Einstein-Hilbert action for
pure gravity,
\begin{equation}
\label{s}
S_G = \frac{1}{2\kappa} \int d^{(1+D)} x \sqrt{-g}R,
\end{equation}
and using the Hawking-Ellis action of a perfect fluid
for the model Universe with variable space dimension the following
Lagrangian has been obtained \cite{r102}
\begin{equation}
\label{34}
L_I := -\frac{V_D}{2 \kappa N} \left( \frac{a}{a_0} \right)
^D D(D-1)
\left[ \left( \frac{\dot a}{a} \right )^2 -\frac{N^2 k}{a^2} \right ]
- \rho N V_D \left( \frac{a}{a_0} \right )^D,
\end{equation}
where $\kappa=8 \pi {M_P}^{-2}=8 \pi G$, $\rho$ the energy density,
and $V_D$ the volume of the space-like sections

\begin{eqnarray}
\label{v1}
V_D&=&\frac{2 \pi^{(D+1)/2}}{\Gamma[(D+1)/2]},\;\;\mbox{closed Universe, $k=+1$,}\\
\label{v2}
V_D&=&\frac{\pi^{(D/2)}}{\Gamma(D/2+1)}{\chi_c}^D,\;\;\mbox{flat Universe, $k=0$,}\\
\label{v3}
V_D&=&\frac{2\pi^{(D/2)}}{\Gamma(D/2)}f(\chi_c),\;\;\mbox{open Universe, $k=-1$.}
\end{eqnarray}

Here $\chi_C$ is a cut-off and $f(\chi_c)$ is a function thereof
(see Ref. \cite{r102}).

Two appropriate questions are raised here.
The first question is that the cut off in Eqs.
(\ref{v2}) and (\ref{v3}) seems {\it ad hoc}, what determines this scale
$\chi_c$? and the second question is that is the boundary term in the
action of Eq. (\ref{s}), more involved with arbitrary dimension?
The answer to the first question is that in cosmology the proper or
physical length are obtained from the comoving length by multiplication of
the Friedmann scale factor $\ell_{\rm physical}=a(t) \ell_{\rm comoving}$.
While the comoving length does not change with time, the proper length
changes with time because of $a(t)$. We take the scale factor having the
dimension of length and the comoving length is a dimensionless quantity.
The comoving length is measured by a set of constant rulers, while the
proper length is measured by a set of expanding or contracting rulers.
The flat model, as we know it is correct for our Universe,
is unbounded with infinite volume, and with infinite radius.
So as explained in Appendix A of Ref. \cite{r102} for flat and open
Universe the volume of the space-like sections are infinite and $\chi_c$
is used as a very large number so that we have 
$$V_D \propto \int^{\chi_c}_0 \chi^{D-1}d\chi,$$
for flat Universe with $k=0$ and
$$V_D \propto \int^{\chi_c}_0 \sinh^{D-1}\chi d\chi,$$
for open Universe with $k=-1$.
The answer to the second question about the boundary term in the action of
Eq. (\ref{s}) is that in the original and substantial paper of \cite{r102}
about this unorthodox model we did not considered this boundary term
since by our knowledge this term does not have crucial effects on the
dynamics of the model. For this reason we ignored the boundary term in the
action of Eq. (\ref{s}) in our original paper of \cite{r102}. One another
answer to the second question is that based on the gravitational action
as given in Weinberg's book \cite{wein} we can ignore the boundary term in
the action of Eq. (\ref{s}).

In the limit of constant space dimensions, or $D=D_0$,
$L_I$ approaches to the Einstein-Hilbert Lagrangian
which is
\begin{equation}
\label{37}
L_{I}^0 := - \frac{V_{D_0}}{2 \kappa_0 N}
\left( \frac{a}{a_0} \right)^{D_0} D_0(D_0-1)
\left[ \left( \frac{\dot{a}}{a} \right)^2 - \frac{N^2 k}{a^2} \right ]
- \rho N V_{D_0} \left( \frac{a}{a_0} \right )^{D_0},
\end{equation}
where $\kappa_0=8\pi G_0$ and the zero subscript in $G_0$ denotes its
present value. So, Lagrangian $L_I$ cannot abandon Einstein's gravity.
Varying the Lagrangian $L_I$ with respect to $N$ and $a$, we find the
following equations of motion in the gauge $N=1$, respectively
\begin{eqnarray}
\label{38}
&&\left( \frac{\dot a}{a} \right)^2 +\frac{k}{a^2} =
\frac{2 \kappa \rho}{D(D-1)},\\
\label{39}
&&(D-1) \bigg\{ \frac{\ddot{a}}{a} + \left[ \left( \frac{\dot a}{a}
\right)^2
+\frac {k}{a^2} \right] \bigg( -\frac{D^2}{2C} \frac{d \ln V_D}{dD}
-1-\frac{D(2D-1)}{2C(D-1)} 
+\frac{D^2}{2D_0} \bigg) \bigg\} \nonumber\\
&&+ \kappa p \bigg( -\frac{d \ln V_D}{dD} \frac{D}{C} 
-\frac{D}{C} \ln \frac{a}{a_0} +1 \bigg) =0.
\end{eqnarray}
Using (\ref{33}) and (\ref{38}), the evolution equation of the space
dimension can be obtained by
\begin{equation}
\label{310}
{\dot{D}}^2= \frac{D^4}{C^2} \left[ \frac{2 \kappa \rho}{D(D-1)}
-k {\delta}^{-2} e^{-2C/D} \right].
\end{equation}
The continuity equation of the model Universe with variable space
dimension
can be obtained by (\ref{38}) and (\ref{39})
\begin{equation}
\label{311}
\frac{d}{dt} \left[ \rho \left( \frac{a}{a_0} \right)^D V_D \right]
+ p \frac{d}{dt} \left[ \left( \frac{a}{a_0} \right )^D V_D \right] =0.
\end{equation}

\subsection{What was considered erroneously in Ref. \cite{r103}?}

As we will show in the next section, there are the dramatic differences
between the conclusions of this paper and our earlier paper
(Ref. \cite{r103}) which also considered time variation of Newton's
constant. Let us now describe a more detailed statement of precisely
what was wrong with Ref. \cite{r103}?

\begin{table}
\caption{Values of $D_P$, $C$ and
$\beta_0$, see Ref.\cite{r103}.}
\begin{tabular}{ccc} \\ \hline\hline
$D_P$         & $C$         & $\beta_0$ \\ \hline\hline
$3$           & $+\infty$   & $0$ \\ \hline
$4$           & $1678.797$  & $-0.004$ \\ \hline
$10$          & $599.571$   & $-0.012$ \\ \hline                             
$25$          & $476.931$   & $-0.015$ \\ \hline
$+\infty$     & $419.699$   & $-0.017$ \\ \hline\hline
\end{tabular}
\end{table}

In \cite{r103}, we generalized a formulation of a one parameter test
theory - in which $\beta$ is a constant parameter - for the time variation
of Newton's constant to the cases where $\beta$ is not constant but a
function of time
\begin{equation}
\label{312}
G=G_0 \left( \frac{t}{t_0} \right)^{\beta(t)}.
\end{equation}
The time derivative of this equation yields
\begin{equation}
\label{313}
\frac{\dot G}{G} = \frac{\beta(t)}{t} + \dot{\beta}(t)
\ln \left( \frac{t}{t_0} \right) \simeq \frac{\beta(t)}{t},
\end{equation}
where $t_0 \simeq 10^{17} \sec$ is the present time and $G_0$ is the
present value of Newton's constant. It is worth mentioning that $\beta(t)$
and its time derivative ${\dot \beta}(t)$  satisfy the following condition
\begin{equation}
\label{314}
\bigg| \frac{{\dot \beta}(t)}{\beta(t)} \bigg| \ll
\bigg| \frac{1}{t \ln \frac{t}{t_0} } \bigg|.
\end{equation}

This condition may not always be valid. Therefore it must be checked for
each case.
Data from big bang nucleosynthesis yields for the present value
of constant $\beta$-parameter
\begin{equation}
\label{315}
|\beta_0| \leq 0.01.
\end{equation}

Comparing the coefficients of $(\dot a/a)^2$ in $L_I$ and $L_I^0$,
as given in Eq.(\ref{34}) and (\ref{37}) respectively, we obtain
\begin{equation}
\label{316}
\frac{G}{V_D D(D-1)}=\frac{G_0}{V_{D_0}D_0(D_0-1)}.
\end{equation}
Time derivative of this equation with assuming a closed Friemann Universe
with $k=+1$ yields the present value of $\beta$ for different
values of $D_P$ and $C$, see Table 2. Comparing the values of $\beta_0$
from Table 2 with observational data, as given by (\ref{315}),
one can rule TVSD models out with $D_P \geq 10$.

There are some criticisms to above treatments as given in
\cite{r103}. Firstly why we considered a closed Universe
while based on recent observational data we know our real
Universe is flat. Secondly our approach in \cite{r103} is
the comparison of the coefficient of $({\dot a}/{a})^2$ in Lagrangians
$L_I$ and $L_I^0$. This comparison does not have reasonable credibility.
In other words, in \cite{r103} about the effective time variation
of Newton's constant in TVSD model the effects of the surface area of the
unit sphere in $D$-space dimensions had not been included and the effects
of the volume of the extra spatial dimensions had been included for closed
Universe which is wrong because we know by recent observatinal data that
our Universe is flat. In this paper our treatment for the time variation
of Newton's constant is based on a credible approach as explained
in the next section.

\section{Time Variation of Newton's Constant in TVSD Theory}

\subsection{Newton's constant in constant $D$-space dimension}

Taking the space dimension is constant and has an arbitrary value,
here we derive the exact relationship between Newton's gravitational
constants $G_{(1+D)}$ and $G_{(1+3)}$, in $(1+D)$ and in $(1+3)$
dimensional theories.

Using the force laws in $(1+D)$ and $(1+3)$ dimensional theories,
which are defined by
\begin{eqnarray}
\label{e31s1}
F_{(1+D)}(r) &=& G_{(1+D)} \frac{m_1 m_2}{r^{D-1}},\\
\label{e31s2}
F_{(1+3)}(r) &=& G_{(1+3)} \frac{m_1 m_2}{r^2},
\end{eqnarray}
and the $(1+D)$ dimensional Gauss' law, one can derive the exact
relationship between the gravitational constants $G_{(1+D)}$, $G_{(1+3)}$
of the full $(1+D)$ and compactified $(1+3)$ dimensional theories 
( see Ref. \cite{r301} for a more detailed explanation)
\begin{equation}
\label{e31s3}
G_{(1+3)} = \frac{S_D}{4 \pi} \frac{G_{(1+D)}}{V_{(D-3)}},
\end{equation}
where $S_D=2 \pi^{D/2}/ \Gamma(D/2)$ is the surface area of the unit
sphere in $D$ spatial dimensions and $V_{(D-3)}$ is the volume of $(D-3)$
extra spatial dimensions.

According to Eq. (\ref{e31s3}),
in a Universe with constant $D$-space Newton's constant
does not have any time variation because all quantities in Eq.
(\ref{e31s3}) are constant.

\subsection{Time variation of the effective $G_{(1+3)}$ from time
variable space dimension}

Let us now generalize Eq. (\ref{e31s3}) from constant $D$-space
dimension to time-varying $D(t)$-space dimension.
We use the following substitution

$$
D={\rm constant}\;\;\to\;\; D(t):=D_t,
$$
where the $t$ subscript in $D_t$ means that $D$ to be as a function
of time. In previous references \cite{r101}-\cite{r107} the notation
of $D_t$ did not use for $D(t)$. Here the author use this notation to
make more clear for the readers.

Therefore, the force law in TVSD theory reads
\begin{equation}
\label{e32s2}
F_{(1+D_t)}(r) = G_{(1+D_t)} \frac{m_1 m_2}{r^{D_t-1}},
\end{equation}
and the exact relationship between Newton's constants
in TVSD theory is

\begin{equation}
\label{e32s3}
G_{(1+3)} = \frac{S_{D_t}}{4 \pi} \frac{G_{(1+{D_t})}}{V_{({D_t}-3)}},
\end{equation}
where $S_{D_t}$ is the surface area of the unit sphere in $D_t$
spatial dimensions
\begin{equation}
\label{e32s4}
S_{D_t}=\frac{2 \pi^{D_t/2}}{\Gamma(D_t/2)},
\end{equation}
and $V_{(D_t-3)}$ is the volume of $(D_t-3)$ extra spatial dimensions
in TVSD theory
\begin{equation}
\label{e32s5}
V_{(D_t-3)} \approx a^{D_t-3},
\end{equation}
where $a$ is the scale factor, see the metric in Eq. (\ref{metric}).

It is easy to see that $G_{(1+D_t)}$, has variable dimension
[length]$^{D_t-1}$ and $G_{(1+3)}$ has constant dimension [length]$^{2}$.
Each time we have varying constants, it is very important to talk about
dimensionless constant such as the fine structure constant
(see Ref. \cite{r304}). In the case of the
Newton's constant, what is meant is usually $G_{(1+3)} m^2$
(more exactly $G_{(1+3)} m^2/\hbar c$) where $m$ is some mass.
So, we introduce a fixed mass scale $M$ that can be taken to be
the Planck mass or some other fixed mass scale.

We take the fixed mass scale to be the Planck mass. 
Therefore, Eq. (\ref{e32s3}) takes the form
\begin{equation}
\label{e32s7}
G_{(1+3)}=\frac{S_{D_t}}{4 \pi} \frac{G_{(1+D_t)}
M_P^{D_t-3}}{(M_P a)^{D_t-3}}.
\end{equation}
The quantity $G_{(1+D_t)}M_P^{D_t-3}$ has dimension [length]$^2$
and we define it
\begin{equation}
\label{e32s8}
{\bar G}_{(1+D_t)} \equiv G_{(1+D_t)}M_P^{D_t-3}.
\end{equation}

It is worth mentioning that if we did not introduce a fixed mass scale
the time variation of Eq. (\ref{e32s3}) runs into serious mathematical
difficulties e.g. $\ln(a)$  which is meaningless because the scale factor
$a$ has dimension [length].

One general feature of extra-dimensional theories, such as
Kaluza-Klein and string theories, is that the ``true'' constants
of nature are defined in the full higher dimensional theory
so that the effective $4$-dimensional constants depends,
among other things, on the structure and size of the extra-dimensions.
Any evolution of these sizes either in time or space, would lead
to a spacetime dependence of the effective $4$-dimensional constants,
see Ref. \cite{r304}.
So ${\bar G}_{(1+D_t)}$ is a ``true'' constant and we have
$\dot{\bar G}_{(1+D_t)}=0$.

Time derivative of Eq. (\ref{e32s7}) takes the form
\begin{equation}
\label{e32s9}
\frac{\dot G_{(1+3)}}{G_{(1+3)}}=\frac{\dot S_{D_t}}{S_{D_t}}+
\frac{\frac{d}{dt}
\bigg[(M_P a)^{D_t-3}\bigg]}
{\bigg[ (M_P a)^{D_t-3}\bigg]}.
\end{equation}
From Eqs. (\ref{31}) and (\ref{33}), we have
\begin{eqnarray}
\label{e32s10}
a&=&\delta e^{C/D_t},\\
\label{e32s11}
\delta&=&a_0 e^{-C/D_0}.
\end{eqnarray}
Using these equations and Eq. (\ref{e32s4}), we get
\begin{equation}
\label{e32s12}
\frac{\dot G_{(1+3)}}{G_{(1+3)}}=
{\dot D_t}\bigg[ \frac{1}{2}\ln \pi -
\frac{1}{2}\psi({D_t}/{2}) -\frac{C}{D_P}+\frac{C}{D_t} \bigg]
+(D_t-3)\frac{\dot a}{a},
\end{equation}
where Euler's psi function $\psi$ is the logarithmic derivative of the
gamma function $\psi(x)\equiv \Gamma'(x)/\Gamma (x)$.
Finally, using Eq. (\ref{33}) we can rewrite
Eq. (\ref{e32s12}) in the form
\begin{equation}
\label{e32s13}
\frac{\dot G_{(1+3)}}{{G}_{(1+3)}}=
-\frac{D^2_t}{C}\frac{\dot a}{a} \bigg[ \frac{1}{2}\ln \pi -
\frac{1}{2}\psi({D_t}/{2}) -\frac{C}{D_P}+\frac{C}{D_t} \bigg]
+(D_t-3)\frac{\dot a}{a}.
\end{equation}

In the limit of constant $3$-space dimension, $C \to +\infty$,
and $D_t=D_P=3$, Eq. (\ref{e32s13}) leads to the time variarion of
Newton's constant today 
\begin{equation}
\label{e32s14}
\frac{\dot G_{(1+3)}}{G_{(1+3)}}
\bigg|_0 = 0.
\end{equation}
This means that Newton's constant
today must be a constant if the space dimension does not change with time.
Using Eq. (\ref{e32s13}) and $H^{-1}_0= 3000 h^{-1}_0 {\rm Mpc}
= 9.2503 \times 10^{27} h^{-1}_0{\rm cm}$ corresponding to the Hubble
constant today
\begin{equation}
\label{e32s15}
H_0= 3.2409 \times 10^{-18} h_0 \sec^{-1}=
1.0227 \times 10^{-10} h_0 {\rm yr}^{-1},
\end{equation}
TVSD theory predicts a decrease in the present value of $G_{(1+3)}$
during the time and derives the time variation of Newton's constant
today when $D_P=4,\,10,\,25,\,+\infty$.
As shown in Table 3, the effect
of time-variable space dimension is that the absolute value of
$\dot G_{(1+3)}/G_{(1+3)}$ to be bigger than that of in
the constant $3$-space dimension. In other words, in constant
$3$-space dimension Newton's constant changes with time less than
in the case of time variable space dimension.

\begin{table}
\caption{Values of $D_P$, $C$ and
${\dot G_{(1+3)}}/G_{(1+3)}|_0$ in terms of sec$^{-1}$ and
yr$^{-1}$.}
\begin{tabular}{cccc}\\ \hline\hline 
$D_P$         & $C$         & ${\dot G_{(1+3)}}/G_{(1+3)}|_0\,({\rm sec}^{-1})$
& ${\dot G_{(1+3)}}/G_{(1+3)}|_0\,({\rm yr}^{-1})$ \\ \hline\hline
$3$           & $+\infty$   & $0$ & $0$ \\ \hline
$4$           & $1678.797$  & $-2.4403\times 10^{-18} h_0$ & $-7.7006\times 10^{-11} h_0$ \\ \hline
$10$          & $599.571$   & $-6.8328\times 10^{-18} h_0$ & $-2.1562 \times 10^{-10}h_0$ \\ \hline                             
$25$          & $476.931$   & $-8.5899 \times 10^{-18} h_0$ & $-2.7106 \times 10^{-10} h_0$ \\ \hline
$+\infty$     & $419.699$   & $-9.7612 \times 10^{-18} h_0$ & $-3.0802 \times 10^{-10} h_0$ \\ \hline\hline 
\end{tabular}
\end{table}

\subsection{Observational limits on the time variation of the space
dimension}

According to Ref. \cite{r304}, the time variation of the gravitational
constant today has an upper limit
\begin{eqnarray}
\label{e32s16}
|\frac{\dot G_{(1+3)}}{G_{(1+3)}}| \bigg|_0
&<& 9 \times 10^{-12} {\rm yr}^{-1}\nonumber\\
&<& 3 \times 10^{-19} {\rm sec}^{-1}.
\end{eqnarray}
Using Eq. (\ref{e32s12}) we get the time variation of the space
dimension today 
\begin{eqnarray}
\dot D_t \bigg|_0 &>& \frac{-9 \times 10^{-12}-(D_0-3)H_0}
{\frac{1}{2} \ln \pi - \frac{1}{2} \psi(D_0/2) +\ln (a_0/\ell_P)}
{\rm yr}^{-1}\nonumber\\
&>&\frac{-3 \times 10^{-19}-(D_0-3)H_0}
{\frac{1}{2} \ln \pi - \frac{1}{2} \psi(D_0/2) +\ln (a_0/\ell_P)}
{\rm sec}^{-1}.
\end{eqnarray}
Taking $D_0=3$ we get a lower limit on the time variation of the
space dimension today
\begin{eqnarray}
\dot D_t \bigg|_0 &>& -6 \times 10^{-14} {\rm yr}^{-1}\\
&>& -2 \times 10^{-21} {\rm sec}^{-1}.
\end{eqnarray}

These values corresponds to a lower limit for the $C$-parameter
in the theory $C> 14365$.
Substituting this value of $C$ in Eq. (\ref{32})
and taking $a=\ell_P$ we get for the space dimension at the Planck length
to be $D_P \leq 3.09$.
This means that according to the limits on the time variation of
the $(1+3)$-dimensional Newton's constant, when the scale factor of the
Universe is equal to the Planck length, the space dimension of
the Universe must be equal to or less than $3.09$.

Let us now consider the data from Cassini spacecraft for the
post-Newtonian parameter $\gamma$ \cite{bert}
\begin{equation}
\label{gamma}
\gamma=1+(2.1\pm 2.3)\times 10^{-5}.
\end{equation}
Cassini's experiment has a too short duration and there is no way one
could deduce anything about the cosmological time variation of Newton's
constant. Also, the uncertainty in the spacecraft state vector
(a few kilometers) is too big to allow a good determination of the
spacecraft longitude. Cassini's data \cite{bert} had
nothing to do with the time variation of Newton's constant.
For the time variation of Newton's constant one needs
much longer experiments, like in the Lunar Laser Ranging experiment.
Furthermore, the value of $\gamma$, as given in Eq. (\ref{gamma}),
can give a constraint on the time variation of Newton's constant, but
this constraint will be model-dependent \cite{will}.
For instance, if we assume a Brans-Dicke theory then
$\gamma=(1+\omega)/(2+\omega)$ and Newton's constant will vary as
$t^{-n}$ with $n^{-1}=2+\frac{3\omega}{2}$. If we use $\gamma$ from
Eq. (\ref{gamma}), we get that $|\omega| > 50000$, so that
$|\frac{\dot G}{G}|_0<\frac{n}{t_{\rm universe}} \sim
\frac{\frac{1}{75000}}{15\times 10^{9} {\rm yr}}\sim 9
\times 10^{-16} {\rm yr}^{-1}$.
Indeed this is $10000$ times smaller than previous bound on time variation
of Newton's constant from Ref. \cite{r304}, see Eq. (\ref{e32s16}).
To get the time variation of the space dimension today based on
Cassini's data,the post-Newtonian parameter $\gamma$ must be calculated
in terms of the parameters of TVSD model. Then one can obtain
${\dot G}/G$ today in the model.
This issue, like calculations for Brans-Dicke
theory in Will's book \cite{will}, needs more mathematical treatments
and is beyond the aim of this article.

\section{Reasons for Ruling the Model Out}
This paper presents a new calculation in the context of an unorthodox
model in which the dimensionality of space is allowed to vary in a
particular prescribed fashion. The model itself is novel but contains
some {\it ad hoc} assumptions. Aside from the mathematical issues
associated with what is meant by spacetimes of non-integer dimensionality,
the particular manner in which variation of this dimensionality is
prescribed - from Eqs. (\ref{31}), (\ref{32}) and (\ref{33}) - is an
{\it ad hoc} assumption which is justified by appealing to polymer physics.
Assumptions which work well for polymers, which themselves can be ascribed
various fractional dimensions while they fill an existing space of fixed
dimension, do not necessarily translate to the dimensionality of space
itself. We can imagine that one could conceive of many other ways in which the
dimension of space could be varied dynamically; and hopefully some of the
ways would have more direct physical motivations coming from quantum
gravity itself, rather than an arbitrary analogy to another part of
physics which may or may not be actually relevant.

Nonetheless, one can sometimes make progress with toy models based on
ad hoc assumptions. For example, Ref. \cite{r301} (Arkani-Hamed
{\it et al.}) makes the {\it ad hoc} assumption that the effective
Planck scale should vary with a distance scale, space becoming
effectively higher-dimensional at very short distances. The model that
we have studied in Refs. \cite{{r102},{r103},{r104},{r105}} makes
a rather different, but equally {\it ad hoc} assumption, namely that there
is a particular Planck scale ${\ell_P}=1.6160 \times 10^{-33} {\rm cm}$
which is absolutely fixed, and the dimension of space is then prescribed
to vary according to Eq. (\ref{31}), varying from some initial dimension
$D_P$, when the scale factor was $\ell_P=1.6160 \times 10^{-33} {\rm cm}$.

Although these assumptions appear to be rather artificial and not based on
any particular natural grounds one might expect in quantum gravity, they
do nonetheless provide the basis for a novel toy model, which has been
extensively studied in Refs. \cite{{r102},{r103},{r104},{r105}}.

It would appear to the author of this paper to be time to ask the
question, can such models be ruled out observationally?
This paper provides a {\it prima facie} case for doing so.
In particular, the conclusion of this paper
is that based on observational bounds on the present-day variation
of Newton's constant, one would have to conclude that $D_P \leq 3.09$.
If the dimension of space when the Universe was ``at the Planck scale''
is constrained to be fractional and very close to $3$, then the whole
edifice of this model loses credibility.

The original hope in this model was that the Universe might be described
by some higher-dimensional unified theory, (e.g., 11-dimensional
$M$-theory etc), and that ``decrumpling'' would provide a dynamical
alternative to compactification in the usual Kaluza-Klein sense. It is
important to study such alternatives since we are really lacking any good
understanding of the ``compactification transition'', if such a process
ever occurred. However, if the result of the calculations is that
$D_P \leq 3.09$ then there can be no place for this alternative to
compactification in the context of a higher-dimensional theory of gravity,
and this particular means of varying the dimension of space dynamically
would appear to be effectively ruled out.

These results are perhaps disappointing, and one might ask what
assumptions might be relaxed or altered. The constraint that $D_0=3$
at the present epoch seems unavoidable. As discussed, e.g. in section
4.8 of Barrow and Tipler \cite{anthropic}, one runs into numerous
problems if one abandons having three dimensions today: waves do not
propagate cleanly, orbits are unstable due to loss of inverse square
law etc. On the other hand, the assumption that the dimensions change at
the constant rate prescribed in Refs.\cite{{r102},{r103},{r104},{r105}},
is difficult to justify on physical grounds. One might expect that during
the early Universe the dimension rapidly altered and then stabilized.
However, one then enters the realm of completely quantum gravity.
The aim of our original model was to explore the possibility that the
effects of a dynamically changing spatial dimension may still have
consequences beyond the very early Universe. This remains possible if we
prescribe the rate of change of spacetime dimensionality to be different
to that considered here and in Refs.\cite{{r102},{r103},{r104},{r105}}.
However, it would be best to have a natural means of prescribing this rate
of change, rather than a purely {\it ad hoc one}.

\section{Conclusions}

The idea put forward in Ref. \cite{r101} that the Universe has a
decrumpling expansion and its spatial dimensions decrease with time
is a bold one but we believe that daring and speculative ideas like
this should be explored.

Recently, the variability of the physical constants for example
fine structure constant, Newton's constant and the speed of light
have attracted much attetion. In this article in the framework of
time variable space dimension (TVSD) theory we have studied the
time variation of ($1+3$)-dimensional Newton's constant. Our treatments
to study this topic is based on the relationship between the
($1+3$) and ($1+D$)-dimensional Newton's constants as obtaind by Gauss'
law and on the ``true'' Newton's constant in the full higher dimensional
theory. Previous studies in Ref. \cite{r103} about the effective time
variation of ($1+3$)-dimensional Newton's constant in TVSD theory had not
been included the effects of the volume of the extra spatial dimensions
and the surface area of the unit sphere in $D$-space dimension. 

Our main result is that the absolute value of the change of the spatial
dimensions must be less than about $10^{-14} {\rm yr}^{-1}$.
This value corresponds to a lower limit for the $C$-parameter
in the theory $C> 14365$.
Substituting this value of $C$ in Eq. (\ref{32})
and taking $a=\ell_P$ we get for the spatial dimension at the Planck
length to be $D_P \leq 3.09$.
This means that according to the limits on the time variation of
the $(1+3)$-dimensional Newton's constant, when the scale factor of the
Universe is equal to the Planck length, the space dimension of
the Universe must be equal to or less than $3.09$.

Based on the time variability of the size of
the extra spatial dimensions Barrow in his study \cite{john} has reported
the present rate of change of the mean radius of any additional spatial
dimensions to be less than about $10^{-19} {\rm yr}^{-1}$.
It is worth mentioning that Barrow's study is based on the dynamical
behavior of the size of extra spatial dimensions while in TVSD theory we
take the size of extra spatial dimensions to be constant and the number of
the spatial dimensions changes with time. 

Indeed, there are certainly many deep physical issues to be explored
in the context of TVSD theory. The question of quantum mechanical
generation of perturbation and their subsequent evolution is of utmost
importance. One can comment upon these issues 
also the WMAP results and the spectral tilt in the context of TVSD model.
One another important question is that what
implications are there for Planck epoch if the spatial dimension at
the Planck length (when the scale factor to be $\ell_P$) to be $3.02$,
or what implications are there for primordial nucleosynthesis
$z \simeq 10^{10}$ if the
spatial dimension at the nucleosynthesis epoch is about $3.13$ as
studied in Ref. \cite{red}. Can experimental observations of light
element abundances be used to rule out any of these models?
Similar questions would apply at all other particular redshift
of cosmological significance, e.g. at recombination $z \simeq 1000$ etc.
It would also be interesting to study a possible variation of the
fine structure constant in the context of TVSD theories.

Our results show, however, that some of the fundamental assumptions of
the TVSD model, as developed in Refs.\cite{{r102},{r103},{r104},{r105}},
need to be altered before these interesting physical questions could be
addressed. While it is common to make {\it ad hoc} assumptions in
cosmological model building in the absence of a complete theory of
quantum gravity, some of the particular ingredients which we have assumed
owe their physical basis perhaps more to polymer physics than to
cosmology. The prescribed rate of change of the spatial dimension,
which is crucial to making predictions with the model, is particularly
hard to justify physically. It is quite possible that this part of the
model should be revised. However, just how this should be done is far from
obvious.

In conclusion, we have shown that the TVSD model of
Refs.\cite{{r102},{r103},{r104},{r105}}, with the constraint that at the
present epoch $D_0=3$ further constraints the space dimension to be
$D\leq 3.09$ at the Planck epoch. This would appear to eliminate the
original motivation of the TVSD model, which was to integrate a variable
space dimension with $M$-theory as an alternative to compactification,
assuming that the spatial dimension was higher at the Planck epoch.
Progress with the TVSD model can only be made if there is a breakthrough
in terms of finding a natural mechanism for varying the spatial dimension
in some alternative fashion to that which we have considered.

\section*{Acknowledgments}

It is a pleasure to thank Jean-Philippe Uzan,
Bruno Bertotti, Luciano Iess, Clifford Will,
Michael Murphy and Josh Frieman for useful discussions and comments.
The author thanks Amir and Shahrokh for useful helps.


\begin{thebibliography}{99}

\bibitem{r101} M. Khorrami, R. Mansouri and M. Mohazzab, Helv. Phys.
Acta {\bf 69}, 237 (1996), {\tt gr-qc/9607049}; M. Khorrami, R. Mansouri,
M. Mohazzab and M. R. Ejtehadi, ``A model Universe with variable
dimension: Expansion as decrumpling'', {\tt gr-qc/9507059}.

\bibitem{r102} R. Mansouri and F. Nasseri, Phys. Rev. D {\bf 60}, 123512
(1999),{\tt gr-qc/9902043}.

\bibitem{r103} R.Mansouri, F.Nasseri and M.Khorrami, Phys.Lett.A
{\bf 259}, 194 (1999), {\tt gr-qc/9905052}.

\bibitem{r104} F. Nasseri and S. Rahvar, Int. J. Mod. Phys. D {\bf 11},
511 (2002), {\tt gr-qc/0008044}; F. Nasseri and S. A. Alavi,
``Noncommutative decrumpling inflation and running of the spectral
index'', {\tt hep-th/0410259}, to appear in Int. J. Mod. Phys. D.

\bibitem{r105} F. Nasseri, Phys. Lett. B {\bf 538},
223 (2002), {\tt gr-qc/0203032}; F. Nasseri and S. Rahvar, ``Dynamics
of Inflationary Cosmology in TVSD Model'', {\tt astro-ph/0212371}.

\bibitem{r106} J. A. S. Lima and M. Mohazzab, Int. J. Mod. Phys. D
{\bf 7}, 657 (1998), {\tt  gr-qc/9607017}.

\bibitem{r107} M. Mohazzab and J. A. S. Lima, Int. J. Mod. Phys. D
{\bf 8}, 751 (1999).

\bibitem{4mn} P. H. Coleman and L. Pietronero, Phys. Rep. {\bf 213},
311 (1992).

\bibitem{5mn} D. N. Schramm, Science {\bf 256}, 513 (1992); H. J. de Vega,
N. S{\'a}nchez, and F. Combes, in {\it Proceedings of the 6th Eric Chalonge
School on Astrofundamental Physics}, edited by N. S{\'a}nchez and A. Zichichi
(Kluwer, Dordrecht, 1998); Astrophys. J. {\bf 500}, 8 (1998).

\bibitem{6mn} M. Tegmark , Class. Quant. Grav. {\bf 14}. L69 (1997);
W. Z. Chao, Phys. Rev. D {\bf 31}, 3079 (1985).

\bibitem{r204} D. Nelson, T. Piran and S. Weinberg, {\it Statistical
Mechanics, Membranes and Surface} (World Scientific, Singapore, 1989).

\bibitem{r205} F. F. Abraham and M. Kardar, Science {\bf 252}, 419 (1991).

\bibitem{r206} P. J. de Gennes, {\it Scaling Concepts in Polymer
Physics} (Cornell University Press, Ithaca, 1973).

\bibitem{r302} J. M. Cline and J. Vinet, Phys. Rev. D {\bf 68},
025015 (2003), {\tt hep-ph/0211284}.

\bibitem{john} J. D. Barrow, Phys. Rev. D {\bf 35}, 1805 (1987).

\bibitem{barrow} J. Yearsley and J. D. Barrow, Class. Quant. Grav.
{\bf 13}, 2693 (1996).

\bibitem{r108} U. Bleyer, M. Mohazzab, and M. Rainer, Astron. Nachr.
{\bf 317}, 3 (1996), {\tt gr-qc/9508035}.

\bibitem{ann} A. M. Green and A. Mazumdar, Phys. Rev. D {\bf 65}, 105022
(2002), {\tt hep-ph/0201209}.

\bibitem{mich} M. T. Anderson, Class. Quant. Grav. {\bf 18}, 5199 (2001),
{\tt gr-qc/0106061}.

\bibitem{tors} T. Bringmann, M. Eriksson and M. Gustafsson,
Phys. Rev. D {\bf 68} (2003) 063516, {\tt astro-ph/0303497}.

\bibitem{gu} Je-An Gu and W-Y.P Hwang, Phys. Rev. D {\bf 66}, 024003
(2002), {\tt astro-ph/0112565}.

\bibitem{r203} M. Gasperini and G. Ummarino, Phys. Lett. B {\bf 266},
275 (1991).

\bibitem{fed} F. Pompilio and M. Montuori, Class. Quant. Grav. {\bf 19}, 203 (2002),
{\tt astro-ph/0111534}.

\bibitem{thd} W. da Cruz,``The Hausdorff Dimension of Fractal Sets and
Fractional Quantum Hall Effects'', Chaos Salitons Fractals, {\bf 17},
975 (2003), {\tt math-ph/0209028}.

\bibitem{tsm} H. J. de Vega and N. S{\'a}nchez, Phys. Lett. B {\bf 490}, 180
(2000), {\tt hep-th/9903236}.

\bibitem{otn} A. E. Allahverdyan, V. G. Gurzadyan and A. A. Soghoyan,
``On the Numerical Study of the Complexity and Fractal Dimension of CMB
Anisotropies'',{\tt astro-ph/9910338}.

\bibitem{hill} C. T. Hill, Phys. Rev. D {\bf 67}, 085004 (2003),
{\tt hep-th/0210076}.

\bibitem{carl} C. Castro, ``On the Four Dimensional Conformal Anomaly,
Fractal Spacetime and the Fine Structure Constant'',
{\tt physics/0010072}.

\bibitem{mittal} A. K. Mittal and D. Lohiya, ``Radiation in a Fractal
Cosmology'', {\tt astro-ph/0104394}; ``From Fractal Cosmography to
Fractal Cosmology'', {\tt astro-ph/0104370}.

\bibitem{lya} L. Ya. Kobelev, ``What Dimensions Do the Time and Space
Have: Integer or Fractional?'', {\tt physics/0001035}.

\bibitem{abda} E. Abdalla, N. Afshordi, K. Khodjasteh and R. Mohayaee,
Astron. $\&$ Astrophys. {\bf 345}, 22 (1999), {\tt astro-ph/9803187}.
\bibitem{r304} J. -P. Uzan, Rev. Mod. Phys. {\bf 75},
403 (2003), {\tt hep-ph/0205340}.

\bibitem{r301} N. Arkani-Hamed, S. Dimopoulos and G. Dvali,
Phys. Rev. D {\bf 59}, 086004 (1999), {\tt hep-ph/9807344}.

\bibitem{sdegl} S. Degl'Innocenti, G. Fiorentini, G. G. Raffelt, B. Ricci
and A. Weiss, Astron. {$\&$} Astrophys. {\bf 312}, 345 (1996),
{\tt astro-ph/9509090}.

\bibitem{fgb} E. Garcia-Berro, Yu. A. Kubyshin and P. Loren-Aguilar,
``Time Variation of the Gravitational and Fine Structure Constants
in Models with Extra Dimensions'', {\tt gr-qc/0302006}, to appear
in the Proceedings of Spanish Relativity Meeting on Gravitation
and Cosmology, ERE-2002 (Mao, Menorca, September 22-24, 2002).

\bibitem{ali} R. Mansouri and A. Nayeri, Grav. $\&$ Cosmol. {\bf 4},
142 (1998), {\tt gr-qc/9609061}.

\bibitem{r201} T. Kaluza, Preus. Akad. Wiss. {\bf K1}, 966 (1921);
O. Klein, Z. Phys. {\bf 37}, 895 (1926); Nature {\bf 118}, 516 (1926).

\bibitem{r202} M. J. Duff, B. E. W. Nilsson and C. N. Pope,
Phys. Rep. {\bf 130}, 1 (1986).

\bibitem{red} F. Nasseri, ``Redshift Dependence of Spatial Dimensions'',
{\tt astro-ph/0212548}.

\bibitem{anthropic} J. D. Barrow and F. J. Tipler, {\it The Anthropic
Cosmological Principle} (Oxford University Press, New York, 1986).

\bibitem{wein} S. Weinberg, {\it Gravitation and Cosmology: Principles
and Applications of the General Theory of Relativity} (John Wiley,
New York, 1972).

\bibitem{bert} B. Bertotti, L. Iess and P. Tortora, Nature {\bf 425} (2003)
374.

\bibitem{will} C. M. Will, {\it Theory and Experiment in Gravitational
Physics} (Cambridge University Press, Cambridge, 1993).

\end{thebibliography}
\end{document}